\documentclass[12pt,english]{article}
\usepackage[a4paper,bindingoffset=0.2in,%
            left=1in,right=1in,top=1in,bottom=1in,%
           ]{geometry}
\usepackage{amsthm}
\usepackage{amssymb}
\usepackage{amsmath}
    \allowdisplaybreaks
\usepackage{autobreak}
\usepackage{amsfonts}
\usepackage{algpseudocode}
\usepackage{chngcntr}
\usepackage[ruled,vlined]{algorithm2e}
\usepackage{algpseudocode}
\usepackage{graphicx}
\usepackage{amstext, graphicx,  latexsym, amscd,  mathtools, amsopn, cite}
\usepackage{subfigure}
\usepackage{mathrsfs}
\usepackage{bm}
\usepackage{subeqnarray,booktabs}
\usepackage{cases}
\usepackage[toc,titletoc]{appendix}
\usepackage{empheq}

\usepackage[sort]{natbib}

\usepackage[normalem]{ulem} 

\usepackage{url}
\usepackage{array}
\newcolumntype{C}{>{\raggedright\arraybackslash}p{1.5cm}<{}}
\usepackage{tabularx}
\usepackage{enumerate}
\usepackage{setspace} 
\usepackage{tikz-cd}
\usepackage{numberedblock}
\usepackage{titlesec}
\titleformat{\section}{\normalsize\bfseries}{\thesection.}{0.5em}{}
\titleformat{\subsection}{\normalsize\bfseries}{\thesubsection.}{0.5em}{}
\usepackage[british]{babel}
\usepackage{indentfirst} 
\usepackage{xcolor}
\usepackage{url}
\usepackage{babel}
\usepackage[utf8x]{inputenc}
\usepackage{array,multirow,graphicx}
\usepackage{float}
\usepackage{placeins}
\usepackage{booktabs}
\usepackage{rotating}
\usepackage{lscape} 

\usepackage{amsmath,booktabs,array}
\usepackage[skip=0.333\baselineskip]{caption}
\newcolumntype{L}{>{$}l<{$}}

\usepackage[bookmarksopen=true]{hyperref}
\usepackage{bookmark}
\usepackage{hyperref, bookmark}
\usepackage{bookmark,etoolbox}
\bookmarksetup{numbered}
\usepackage{hyperref}
\usepackage{amsmath,amsfonts,amssymb,amsthm}
\usepackage{chngcntr}
\theoremstyle{plain}
\newtheorem{assumption}{Assumption}

\renewcommand{\theequation}{\arabic{equation}}
\def\b1{\mbox{\boldmath $1$}}

\hypersetup{
    colorlinks=true,
    linkcolor=blue,
}
 \usepackage{hyperref}
 \hypersetup{
     colorlinks=true,
     linkcolor=palecarmine,
     filecolor=blue,
     citecolor = black,      
     urlcolor=blue,
     }
\definecolor{lava}{rgb}{0.81, 0.06, 0.13}
\definecolor{lust}{rgb}{0.9, 0.13, 0.13}
\definecolor{utahcrimson}{rgb}{0.83, 0.0, 0.25}
\definecolor{red-brown}{rgb}{0.65, 0.16, 0.16}
\definecolor{palecarmine}{rgb}{0.69, 0.25, 0.21}
\definecolor{internationalkleinblue}{rgb}{0.0, 0.18, 0.65}
\definecolor{dukeblue}{rgb}{0.0, 0.0, 0.61}
\definecolor{darkpowderblue}{rgb}{0.0, 0.2, 0.6}
\definecolor{ceruleanblue}{rgb}{0.16, 0.32, 0.75}
\begin{document}
\title{Particle MCMC in forecasting frailty correlated default models with expert opinion}

\author{Ha Nguyen \footnote {Department of Actuarial Studies and Business Analytics,
Macquarie Business School, Macquarie University, Sydney, NSW 2109, AUSTRALIA;
Email: ha.t.nguyen@mq.edu.au, ha.nguyen99804@gmail.com}
}
\date{\vspace{-5ex}} 
\begin{titlepage}
\clearpage
\maketitle\thispagestyle{empty}

\begin{abstract}
\noindent
Predicting corporate default risk has long been a crucial
topic in the finance field, as bankruptcies impose enormous costs on market participants as well as the economy as a whole. This paper aims to forecast frailty correlated default models with subjective judgements on a sample of U.S. public non-financial firms spanning January 1980-June 2019. We consider a reduced-form model and adopt a Bayesian approach coupled with the Particle Markov Chain Monte Carlo (Particle MCMC) algorithm to scrutinize this problem. The findings show that the 1-year prediction for frailty correlated default models with different prior distributions is relatively good, whereas the prediction accuracy ratios for frailty-correlated default models with non-informative and subjective prior distributions over various prediction horizons are not significantly different.
\end{abstract}

\end{titlepage}
\newpage

\section{Introduction}
\vspace{-0.5cm}
Default forecasting is crucial for financial institutions and investors. Prior to investing in or extending credit to a company, investors and creditors must assess the company's financial distress risk in order to avoid incurring a significant loss.  In the literature on financial distress, default risk modelling can be grouped into two main categories: structural and reduced form approaches. This paper uses the reduced-form method of correlated default timing. The interested readers may refer to  \hyperlink{thesentence}{\textcolor{ceruleanblue}{Nguyen and Zhou \textcolor{black}{(}2023\textcolor{black}{)}}} for a general view of the literature on reduced-form models of correlated default timing.

Accounting-based measures are the first generation of reduced-form models for predicting the failure of a company. The earliest works predicting this type of financial distress is univariate analysis (\hyperlink{thesentence}{\textcolor{ceruleanblue}{Beaver, 1966, 1968}}), who employ financial ratios independently and adopts the cut-off point for each financial ratio in order to improve the precision of classifications for a distinct sample. \hyperlink{thesentence}{\textcolor{ceruleanblue}{Altman \textcolor{black}{(}1968\textcolor{black}{)}}} conducted a multivariate analysis of business failure based on multiple discriminant analyses by combining the data from numerous financial ratios from the financial statement into a singular weighted index. The second generation of default literature is the logistic model (\hyperlink{thesentence}{\textcolor{ceruleanblue}{Ohlson, 1980}}). This method was developed to deal with the shortcomings of the Altman Z-score method. \hyperlink{thesentence}{\textcolor{ceruleanblue}{Shumway \textcolor{black}{(}2001\textcolor{black}{)}}} attempts to predict defaults and shows that half of the accounting ratios utilised by \hyperlink{thesentence}{\textcolor{ceruleanblue}{Altman \textcolor{black}{(}1968\textcolor{black}{)}}} and \hyperlink{thesentence}{\textcolor{ceruleanblue}{Zmijewski \textcolor{black}{(}1984\textcolor{black}{)}}} have poor prediction on the default models, while a large number of market-driven independent variables are significantly associated with default probability. The recent expansion of reduced form default risk models has centred on duration analysis. \hyperlink{thesentence}{\textcolor{ceruleanblue}{Jaarrow and Turnbull \textcolor{black}{(}1995\textcolor{black}{)}}} and \hyperlink{thesentence}{\textcolor{ceruleanblue}{Jarrow et al. \textcolor{black}{(}1997\textcolor{black}{)}}} are the pioneers of term structure and credit spread modelling. 

With regard to duration analysis, recent research indicates that observable macroeconomic and firm-specific factors may not be sufficient to characterize the variation in default risk, as corporate default rates are strongly correlated with latent factors. The need for and importance of the hidden factor in a default model are discussed in several recent studies, such as \hyperlink{thesentence}{\textcolor{ceruleanblue}{Koopman and Lucas \textcolor{black}{(}2008\textcolor{black}{)}}}, \hyperlink{thesentence}{\textcolor{ceruleanblue}{Koopman et al. \textcolor{black}{(}2009\textcolor{black}{)}}}, \hyperlink{thesentence}{\textcolor{ceruleanblue}{Duffie et al. \textcolor{black}{(}2009\textcolor{black}{)}}}, \hyperlink{thesentence}{\textcolor{ceruleanblue}{Chava et al. \textcolor{black}{(}2011\textcolor{black}{)}}}, \hyperlink{thesentence}{\textcolor{ceruleanblue}{Koopman et al. \textcolor{black}{(}2011, 2012\textcolor{black}{)}}}, \hyperlink{thesentence}{\textcolor{ceruleanblue}{Creal et al. \textcolor{black}{(}2014\textcolor{black}{)}}}, \hyperlink{thesentence}{\textcolor{ceruleanblue}{Azizpour et al. \textcolor{black}{(}2018\textcolor{black}{)}}}, and \hyperlink{thesentence}{\textcolor{ceruleanblue}{Nguyen \textcolor{black}{(}2023\textcolor{black}{)}}}. 

To improve the prediction accuracy of default models, the utilization of expert judgement in the decision-making process is common in practice, as there may not be enough statistically significant empirical evidence to reliably estimate the parameters of complicated models. This problem is considered to be of central interest in simulating a number of debates in the empirical literature regarding the issue of Bayesian inference. In the process of inference, however, the majority of Bayesian analyses utilise non-informative priors formed by formal principles. The theoretical foundation utilised by the majority of Bayesians is that of \hyperlink{thesentence}{\textcolor{ceruleanblue}{Savage \textcolor{black}{(}1971, 1972\textcolor{black}{)}}}
and \hyperlink{thesentence}{\textcolor{ceruleanblue}{De Finetti \textcolor{black}{(}2017\textcolor{black}{)}}}. Despite the fact that non-informative prior distribution plays a crucial role in defining the model for certain problems, it appears that there is an unavoidable drawback, as it is sometimes impossible to specify only non-informative priors and disregard the informative priors. It is observed that Bayes factors are sensitive to the selection of unknown parameters of informative prior distributions, which has a greater likelihood of influencing the posterior distribution. As a consequence, it generates debates regarding the selection of priors. Moreover, real prior information is beneficial for specific applications, whereas non-informative priors do not take advantage of this; consequently, such circumstances require informative priors. In other words, this is where subjective views and expert opinion are combined. Assuming we have a complex, high-dimensional posterior distribution, it is uncertain whether we have exhaustively summarised it. This should likely be completed by an experienced statistician. Choosing informative priors and establishing a connection with expert opinion are still the subject of debate in academic research, and interesting stories about them are still being continued. Recently, there has been research on the default prediction combined with expert opinion using machine learning techniques, such as by \hyperlink{thesentence}{\textcolor{ceruleanblue}{Lin and McClean \textcolor{black}{(}2001\textcolor{black}{)}}}, \hyperlink{thesentence}{\textcolor{ceruleanblue}{Kim and Han \textcolor{black}{(}2003\textcolor{black}{)}}}, \hyperlink{thesentence}{\textcolor{ceruleanblue}{Zhou et al. \textcolor{black}{(}2015\textcolor{black}{)}}}, and \hyperlink{thesentence}{\textcolor{ceruleanblue}{Gepp et al. \textcolor{black}{(}2018\textcolor{black}{)}}}. However, these studies adopt machine learning techniques with single classifiers.

Motivated by these findings, this paper aims to answer the research question of whether adding expert opinions to the frailty correlated default risk model can give us better prediction results. To do so, we combine prior distributions to the frailty correlated default model in \hyperlink{thesentence}{\textcolor{ceruleanblue}{Duffie et al. \textcolor{black}{(}2009\textcolor{black}{)}}} and adopt the Particle MCMC approach in \hyperlink{thesentence}{\textcolor{ceruleanblue}{Nguyen \textcolor{black}{(}2023\textcolor{black}{)}}} to estimate the unknown parameters and predict the default risk in the model using the dataset of U.S. public non-financial firms spanning 1980-2019. Our findings show that the 1-year prediction for frailty correlated default models with different prior distributions is relatively good, whereas the prediction accuracy of models decrease significantly as the prediction horizons increase. The results also indicate that prediction accuracy ratios for frailty-correlated default models with non-informative and subjective prior distributions over various prediction horizons are not significantly different. Specifically, the out-of-sample prediction accuracy for the frailty correlated default models with subjective prior distribution is slightly higher than that of the frailty correlated default models with uniform prior distribution (95.00\% for 1-year prediction, 85.23\% for 2-year prediction, and 83.18\% for 3-year prediction of the frailty default model with uniform prior distribution; and 96.05\% for 1-year prediction, 86.32\% for 2-year prediction, and 84.71\% for 3-year prediction of the frailty default model with subjective prior distribution). 

To obtain the research objective, the remainder of the paper is organized as follows: Section 2 presents the econometric model and the estimation methodology for the frailty correlated default models with the different prior distributions. Section 3 reports major results. Data and the choice of covariates are also presented in this section. Section 4 provides the model performance evaluation. Section 5 presents the concluding remarks and limitations of the research.
\vspace{-0.5cm}
\section{Econometric Model}
\vspace{-0.5cm}
\renewcommand{\theequation}{3.\arabic{equation}}
\setcounter{equation}{0}
This part outlines the econometric model used by \hyperlink{thesentence}{\textcolor{ceruleanblue}{Duffie et al. \textcolor{black}{(}2009\textcolor{black}{)}}} and our extension to model and improvement of method to examine and forecast default risk at the firm level. We first provide an introduction to the notations used in this study. We consider a complete filtered probability space $(\Omega, {\cal F}, {\cal G}, P)$, where the filtration ${\cal G} := \{ {\cal G}_t \}_{t \in [0, T]}$ describes the flow of information over time and $P$ is a real-world probability measure. Further on, we use the standard convention where capital letters denote random variables, whereas lower case letters are used for their values.

The complete Markov state vector is described as $W_t = (X_{it}, Y_t,H_t)$, where 
$W_t$ be a Markov state vector of firm-specific and macroeconomic covariates; $X_{it}$ a vector of observable firm-specific covariates for firm $i$ at the first observation time $t_i$ until the last observation time $T_i$, $V_i$ be an unobservable firm-specific covariate; $Y_t$ be a vector of observable macroeconomic variables at all times, and $H_t$ be an unobservable frailty (latent macroeconomic factor) variable; $Z_{it} = (1, X_{it}, Y_t)$ denote a vector of observable covariates for firms $i$ at time $t$, where 1 is a constant.

On the event of $s > t$ of survival to $t$, given the information set ${\cal F}_t$, the conditional probability of survival to time $ t + \tau$ is
\begin{equation} \label{survival probability}
q(W_t, \tau) = p(s > t + \tau |{\cal F}_t )= E \bigg(e^{- \int_t^{t + \tau} \lambda(z) dz} |W_t \bigg)
\end{equation}
and the conditional default probability at time $t + \tau$ is of the form:
\begin{equation} \label{default probability}
 p(W_t,\tau) = p(T < t + \tau | {\cal F}_t) =  E\bigg(\int_t^{t+\tau} e^{-\int_t^u\lambda(z)dz} \lambda(u)du|W_t \bigg)
\end{equation}
The information filtration of $\{ {\cal F}_t \}_{t \in [0, T]}$ includes the information set of the observed macroeconomic/firm-specific variables.\\
$$\{Y_\tau \}_{\tau \in [0,t]} \cup \{(X_{i\tau}, D_{i\tau})\}_{i \in [1, m], \tau \in [t_i, min (t, T_i)]}$$
The complete information filtration $\{ {\cal G}_t \}_{t \in [0, T]}$ contains the variables in the information filtration of $\{ {\cal F}_t \}_{t \in [0, T]}$ and the frailty process $\{H_\tau \}_{ \tau \in [0,t]}$.

The assumptions are imposed as follows:
\begin{assumption} \label{As1}
\textit{All firms' default intensities at time $t$ depend on a Markov state vector $W_t$ which is only partially observable}.
\end{assumption}
\begin{assumption} \label{As2}
\textit{Conditional on the path of the Markov state process $W$ determining the default intensities, the firm default times are the first event times of an independent Poisson process with time-varying intensities determined by the path of $W$. This is referred to as a doubly stochastic assumption}. 
\end{assumption}
\begin{assumption} \label{As3}
\textit{ Set the level of mean-reversion of $H$, $\alpha = 0$, the unobserved frailty process $H$ is a mean-reverting Ornstein-Uhlenbeck (OU) process which is given by the stochastic differential equation below:
\begin{equation} \label{OU process}
dH_t = -\eta H_t dt + \sigma dW_t,
\end{equation}
}
where $\eta, \alpha, \sigma$ are parameters; $\{{W_t}\}_{t \in [0,T]}$ is a standard Brownian motion with respect to $(\Omega, {\cal F}, {\cal G}, P)$; $\eta$ is a nonnegative constant, the speed of mean-reversion of $H$; $\sigma$ is the volatility of the Brownian motion.
\end{assumption}
In the general case, without assumption \ref{OU process}, we would need extremely numerically intensive Monte Carlo integration in a high dimensional space due to our large dataset from 1980 to 2019. Thus, we assume process H is an OU process, as in \hyperlink{thesentence}{\textcolor{ceruleanblue}{Duffie et al. \textcolor{black}{(}2009\textcolor{black}{)}}}.

The default intensity of a firm $i$ at the time $t$ is:
 $\lambda_{it} = \Lambda(S_i(W_{t}),\theta)$, where $S_i(W_{t}) = (Z_{it}, H_t)$ is the component of the state vector at time $t$ and $\theta = (\kappa,\xi,\eta,\sigma)$ is a parameter vector to be estimated; $\kappa$ is a parameter vector of the observable covariates $Z$; $\xi$ is a parameter of the frailty variable $H_t$, $\eta$ is the speed of mean-reversion of $H_t$; and $\sigma$ is a Brownian motion parameter of $H_t$. The parameters $\eta$ and $\sigma$ need to be estimated through a  mean-reverting OU process, which we assume the unobserved frailty process $H$ will follow. The proportional hazards form is expressed by
\begin{equation}
\Lambda((z,h), \theta) = e^{(\kappa_1z_1 + \hdots + \kappa_n z_n +  \xi h)}
\end{equation}

$D_m$ is the default indicators of $m$ firms. Default indicator $D_{it}$ of the firm $i$ at the time $t$ is defined as:
\begin{equation*}
  D_{it} =
    \begin{cases}
      1 & \text{if firm $i$ defaulted at time $t$}\\
      0 & \text{otherwise}
    \end{cases}       
\end{equation*}

Now well start with the conditional probability of the $m$ company. As mentioned above, we let $t_i$ be first observation time for firm $i$ and $T_i$ is last observation time for firm $i$. For each firm $i$ and fix the time $t$, we have
$$P(D_{it} = 1|\gamma,\theta) = \lambda_{it}\Delta_t e^{-\lambda_{it}\Delta_t}$$
$$P(D_{it} = 0|\gamma,\theta) = e^{-\lambda_{it}\Delta_t}$$
then, in our case, the conditional probability of the individual firm is given by\\
$$p(Z_{it}, D_{it}, H  |\theta)= \lambda_{it}\Delta_t e^{-\lambda_{it}\Delta_t} D_{it} + e^{-\lambda_{it}\Delta_t} (1- D_{it})$$
\begin{eqnarray} \label{probability with all firms}
{ \prod_{t=t_i}^{T_i}  p(Z_{it}, D_{it}, H  |\theta)}
& = &{e^{-\sum_{t = t_i}^{T_i} {\lambda_{it} \Delta_t}} \prod_{t = t_i}^{T_i}(D_{it} \lambda_{it} \Delta_t + (1 - D_{it}))}
\end{eqnarray}
Thus, the conditional probability of the $m$ firm is expressed as:\\
\begin{eqnarray} \label{probability with all firms}
{\prod_{i = 1}^{m} \bigg(\prod_{t=t_i}^{T_i}  p(Z_{it}, D_{it}, H  |\theta)\bigg)}
& = &{\prod_{i = 1}^{m} \bigg [{e^{-\sum_{t = t_i}^{T_i} {\lambda_{it} \Delta_t}} \prod_{t = t_i}^{T_i}(D_{it} \lambda_{it} \Delta_t + (1 - D_{it}))}\bigg]} \nonumber\\
&=&{e^{-\sum_{i = 1}^{m}\sum_{t = t_i}^{T_i} {\lambda_{it} \Delta_t}}\prod_{i = 1}^{m} \prod_{t = t_i}^{T_i}(D_{it} \lambda_{it} \Delta_t + (1 - D_{it}))}.
\end{eqnarray}
Applying Bayes' theorem,
\begin{equation}
p(\theta|Z,D,H) \quad \propto \quad {\cal L} (\theta|Z, D, H) p(\theta)
\end{equation}
Related to prior distribution, $p(\theta)$, we have 2 cases: (i) Uniform prior and (ii) Subjective prior.
\begin{enumerate}
\item[1.] Prior distribution is uniform 
\begin{equation}
p(\theta|Z,D,H) \quad \propto \quad {\cal L}(Z,D,H|\theta),
\end{equation}
where $p(\theta) \propto 1$ (non-informative prior distribution).
This case is exactly studied by \hyperlink{thesentence}{\textcolor{ceruleanblue}{Duffie et al. \textcolor{black}{(}2009\textcolor{black}{)}}}. Our extension to the model by combining with priors as below
\item[2.] Prior distribution is subjective
\begin{equation}
p(\theta|Z,D,H) \quad \propto \quad {\cal L} (\theta|Z, D, H) {\cal N}(\kappa, \xi|\mu,\Sigma),
\end{equation}
where ${\cal N} (\mu, \Sigma)$ is multivariate normal prior with a mean vector $\mu$ and a covariance matrix $\Sigma$. 
\end{enumerate}
If the observable covariate process $Z$ is independent of the frailty process $H$, the likelihood function of intensity parameter vector $\theta$ is given by
\begin{align} \label{main equation}
{{\cal L} (\theta|Z, D)} 
&= {\int {{\cal L} (\theta| Z, D, h)} p_H(h)dh} \times {\cal N}(\kappa, \xi|\mu,\Sigma)\nonumber \\
&= {E\Bigg [\prod_{i = 1}^{m} \bigg (e^{-\sum_{t = t_i}^{T_i} {\lambda_{it} \Delta_t}} \prod_{t = t_i}^{T_i}(D_{it} \lambda_{it} \Delta_t + (1 - D_{it})) \bigg) \bigg |Z, D \bigg]} \times {\cal N}(\kappa, \xi|\mu,\Sigma),
\end{align}
where $p_H(.)$ is the unconditional probability density of the unobservable frailty process $H$.

Now we show how to transform the model with the frailty correlated defaults to the one combining with subjective prior distribution.
We have found the posterior probability density earlier as
\begin{eqnarray} \label{Posterior}
p(\theta|Z,D,H) &\propto&{\cal L} (\theta|Z, D, H) p(\theta) \nonumber\\
&=&{\cal L} (\theta|Z, D, H) {\cal N}(\kappa, \xi|\mu,\Sigma) \nonumber\\
&=& {\cal L} (\theta|Z, D, H)\frac{ exp(-\frac{1}{2}((\kappa,\xi) - \mu)^T \Sigma^{-1} ((\kappa,\xi) - \mu)}{\sqrt{(2\pi)^n |\Sigma|}}. 
\end{eqnarray}
Taking the logarithm for Eq. (\ref{Posterior})
\begin{equation*} 
log \bigg({\cal L} (\theta|Z, D, H)\frac{ exp(-\frac{1}{2}((\kappa,\xi) - \mu)^T \Sigma^{-1} ((\kappa,\xi) - \mu))}{\sqrt{(2\pi)^n |\Sigma|}} \bigg)	
\end{equation*}
\begin{eqnarray}\label{after taking log}
&=& log ({\cal L} (\theta|Z, D, H) + log \bigg(\frac{ exp(-\frac{1}{2}((\kappa,\xi) - \mu)^T \Sigma^{-1} ((\kappa) - \mu))}{\sqrt{(2\pi)^n |\Sigma|}}\bigg).
\end{eqnarray}
Recall that the log-likelihood of parameter value $\theta$ given the observable and hidden variables is given by
\begin{eqnarray}
{\ell(\theta|Z,D,H)}& = &{log \bigg(e^{-\sum_{i = 1}^{m}\sum_{t = t_i}^{T_i} {\lambda_{it} \Delta_t}}\prod_{i = 1}^{m} \prod_{t = t_i}^{T_i}(D_{it} \lambda_{it} \Delta_t + (1 - D_{it}))\bigg)} \nonumber\\
&=&{-\sum_{i = 1}^{m}\sum_{t = t_i}^{T_i} {\lambda_{it} \Delta_t} + \sum_{i = 1}^{m}\sum_{t = t_i}^{T_i}log(D_{it} \lambda_{it} \Delta_t + (1 - D_{it}))}.
\end{eqnarray}
Now we proceed to take the logarithm for the second term of the Eq.({\ref{after taking log}	)
\begin{equation} \label{take log the 2nd term of the expert opinion equation}
log \bigg(\frac{ exp(-\frac{1}{2}((\kappa,\xi) - \mu)^T \Sigma^{-1} ((\kappa,\xi) - \mu))}{\sqrt{(2\pi)^n |\Sigma|}} \bigg) = - \frac{1}{2\sqrt{(2\pi)^n |\Sigma|}} ((\kappa,\xi) - \mu)^T \Sigma^{-1} ((\kappa,\xi) - \mu).
\end{equation}
In the second term, the central interest is the covariance matrix. For notational simplicity, set $\gamma = (\kappa, \xi)$, it is then rewritten as
\[
\begin{pmatrix}
 (\gamma_1 - \mu_1) c_{1,1} + (\gamma_2 - \mu_2) c_{2,1} + ...+ (\gamma_{n+1} - \mu_{n+1}) c_{n,1}         \\
 (\gamma_2 - \mu_1) c_{1,2} + (\gamma_2 - \mu_2) c_{2,2} + ...+ (\gamma_{n+1} - \mu_{n+1}) c_{n,2}         \\
 \vdots                   \\
 (\gamma_1 - \mu_1) c_{1,n} + (\gamma_2 - \mu_2) c_{2,n} + ...+ (\gamma_{n+1} - \mu_{n+1}) c_{n,n} \\
  (\gamma_1 - \mu_1) c_{1,n+1} + (\gamma_2 - \mu_2) c_{2,n+1} + ...+ (\gamma_{n+1} - \mu_{n+1}) c_{n,n+1} \\ 
 \end{pmatrix}
\begin{pmatrix}
    \gamma_1 - \mu_1         \\
    \gamma_1 - \mu_2         \\
    \vdots                   \\
     \gamma_n- \mu_n          \\
     \gamma_{n+1} - \mu_{n+1} \\
\end{pmatrix}
\]
\begin{eqnarray} \label{transform eq}
&=& (\gamma_1 - \mu_1)\bigg((\gamma_1 - \mu_1)c_{1,1} + (\gamma_2 - \mu_2)c_{2,1} +...+ (\gamma_{n} - \mu_{n})c_{n,1}  +(\gamma_{n+1} - \mu_{n+1}) \nonumber\\
&& c_{n+1,1}  \bigg)  +  (\gamma_2 - \mu_2)\bigg((\gamma_1 - \mu_1)c_{1,2} + (\gamma_2 - \mu_2)c_{2,2} +...+(\gamma_{n} - \mu_{n})c_{n,2} \nonumber\\
&&  + (\gamma_{n+1} - \mu_{n+1})c_{n+1,2} \bigg) + \cdots + (\gamma_{n} - \mu_{n}) \bigg((\gamma_1 - \mu_1)c_{1,n} + (\gamma_2 - \mu_2)c_{2,n} \nonumber\\
&&   +...+(\gamma_{n} -\mu_{n}) c_{n, n}  + (\gamma_{n+1} - \mu_{n+1})c_{n+1,n} \bigg) +(\gamma_{n+1} - \mu_{n+1})\bigg((\gamma_1 - \mu_1)c_{1,n+1} \nonumber\\
&&  + (\gamma_2 - \mu_2)c_{2,n+1} +...+(\gamma_n -\mu_n) c_{n, n+1}  + (\gamma_{n+1} - \mu_{n+1})c_{n+1,n+1} \bigg)\nonumber\\
&=& \bigg ((\gamma_1 -\mu_1)(\gamma_1 - \mu_1) c_{1,1} + (\gamma_1 - \mu_1)(\gamma_2 - \mu_2)c_{2,1} + ...+ (\gamma_1 -\mu_1) (\gamma_n - \mu_n) c_{n,1} \nonumber\\
&& +(\gamma_1 - \mu_1)(\gamma_{n+1} -\mu_{n+1})c_{n+1,1}   \bigg) + \bigg((\gamma_2 -\mu_2)(\gamma_1 - \mu_1) c_{1,2} + (\gamma_2 - \mu_2) \nonumber\\
&&  (\gamma_2 - \mu_2)c_{2,2} + ...+ (\gamma_2 -\mu_2)(\gamma_n - \mu_n) c_{n,2} +(\gamma_2 - \mu_2)(\gamma_{n+1} -\mu_{n+1})c_{n+1,2} \bigg) \nonumber\\
&&+\cdots \nonumber\\
&&  + \bigg ( (\gamma_{n} -\mu_{n})(\gamma_1 - \mu_1) c_{1,n} + (\gamma_n - \mu_n)(\gamma_2 - \mu_2)c_{2,n} + ...+ (\gamma_{n} -\mu_{n}) \nonumber\\
&& (\gamma_n - \mu_n) c_{n,n} +(\gamma_n - \mu_n)(\gamma_{n+1} -\mu_{n+1})c_{n+1,n} \bigg) +\bigg((\gamma_{n+1} -\mu_{n+1})(\gamma_1 - \mu_1)  \nonumber\\
&&c_{1,n+1} + (\gamma_{n+1} - \mu_{n+1})(\gamma_2 - \mu_2)c_{2,n+1} + ... + (\gamma_{n+1} -\mu_{n+1})(\gamma_{n} - \mu_{n})\nonumber\\
&& c_{n,n+1} +(\gamma_{n+1} - \mu_{n+1})(\gamma_{n+1} -\mu_{n+1})c_{n+1,n+1} \bigg) \nonumber\\
&=&\sum_{j=1}^{n+1} \sum_{k=1}^{n+1} c_{jk}(\gamma_j -\mu_j)(\gamma_k -\mu_k).
\end{eqnarray}
Then, the second term can be rewritten as
\begin{equation*}
log \bigg(\frac{ exp(-\frac{1}{2}(\gamma - \mu)^T \Sigma^{-1} (\gamma - \mu))}{\sqrt{(2\pi)^n |\Sigma|}} \bigg) = - \frac{1}{2\sqrt{(2\pi)^n |\Sigma|}} (\gamma - \mu)^T \Sigma^{-1} (\gamma - \mu)
\end{equation*}
\begin{eqnarray}
&=& - \frac{1}{2\sqrt{(2\pi)^n |\Sigma|}}\sum_{j=1}^{n+1} \sum_{k=1}^{n+1} c_{jk}(\gamma_j -\mu_j)(\gamma_k -\mu_k).
\end{eqnarray}
Now, we combine terms of Eq. (\ref{Posterior}) to get an overall likelihood function given the filtration $\cal G$
\begin{equation*}
log ({\cal L} (\theta|Z, D, H) + log \bigg(\frac{ exp(-\frac{1}{2}(\gamma - \mu)^T \Sigma^{-1} (\gamma - \mu))}{\sqrt{(2\pi)^n |\Sigma|}}\bigg)
\end{equation*}
\begin{eqnarray} \label{Expertopinion}
&=& \bigg({-\sum_{i = 1}^{m}\sum_{t = t_i}^{T_i} {\lambda_{it} \Delta_t} + \sum_{i = 1}^{m}\sum_{t = t_i}^{T_i}ln(D_{it} \lambda_{it} \Delta_t + (1 - D_{it}))} \bigg) \nonumber\\
&& + \sum_{j=1}^{n+1} \sum_{k=1}^{n+1} c_{jk}(\gamma_j -\mu_j)(\gamma_k -\mu_k).
\end{eqnarray}

Now the central interest is to estimate Eq. \ref{Expertopinion}. We used a Bayesian approach coupled with the Particle MCMC algorithm to estimate and forecast the frailty correlated default models with uniform and subjective prior distributions. Particle filters can be understood as sequential Monte Carlo (SMC) methods, is introduced by \hyperlink{thesentence}{\textcolor{ceruleanblue}{Handschin and Mayne \textcolor{black}{(}1969\textcolor{black}{)}}} and \hyperlink{thesentence}{\textcolor{ceruleanblue}{Handschin \textcolor{black}{(}1970\textcolor{black}{)}}}. Particles are a set of points in the sample space, and particle filters provide approximation to the posterior densities via these points. Each particle has an assigned weight, and then the posterior distribution can be approximated by a discrete distribution. Several algorithms about particle filters have been proposed in the literature review, and it can be said that the difference between algorithms consist in the way that a set of the particles evolves and adapts to inputs data. Algorithm \ref{SMC algorithm} presents the Sequential Monte Carlo process we applied in our method.
\begin{algorithm}
\caption{Sequential Monte Carlo algorithm} \label{SMC algorithm}
\textsl{•} \text{ At time t = 1}: $\quad \forall n = 1,..., N$\\
\begin{enumerate}
\item[(1)] \textit{Sample} $H_1^n \sim q_{\theta} (.|(z_1, D_1))$
\item[(2)] \textit{Calculate and normalize the weights}
\end{enumerate}
\begin{eqnarray}
w_1(H_1^n) &:= & \frac{p_{\theta} (H_1^n, (z_1, D_1))}{q_{\theta} (H_1^n |(z_1, D_1))} \nonumber\\
&=& \frac{\mu_{\theta} (H_1^n) g_{\theta} ((z_1, D_1)|H_1^n)}{q_{\theta} (H_1^n |(z_1, D_1))}, \nonumber
\end{eqnarray}
$$W_1^n := \frac{w_1 (H_1^n)}{\sum_{i = 1}^N w_1(H_1^i)}.$$
\textsl{•}\text{ At time $t = 2, ..., T$} : $\quad \forall n = 1,..., N$\\
\begin{enumerate}
\item[(1)] \textit{Resample the particles, i.e. sample the indices} $A_{t-1}^n \sim {\cal G}(.|W_{t-1})$,
\item[(2)] \textit{Sample} $H_t^n \sim q(.|((z_t, D_t), H_{t-1}^{A_{t-1}^n}))$
\textit{and set} $H_{1:t}^n := (H_{1:t-1}^{A_{t-1}^n}, H_{t}^n),$
\item[(3)] \textit{Calculate and normalize the weights}
\begin{eqnarray} \label{SMC compute weights}
w_t(H_{1:t}^n) &:=& \frac{p_{\theta}(H_{1:t}^n, (z_{1:t}, D_{1:t}))}{p_{\theta} (H_{1:t-1}^{A_{t-1}^n}, (z_{1:t-1}, D_{1:t-1})) q_{\theta}(H_t^n|((z_t, D_t), H_{t-1}^{A_{t-1}^n}))} \nonumber\\
&=& \frac{f_{\theta} (H_t^n|H_{t-1}^{A_{t-1}^n}) g_{\theta} ((z_t,D_t)|H_t^n)}{q_{\theta} (H_t^n|((z_t, D_t), H_{t-1}^{A_{t-1}^n}))} \nonumber
\end{eqnarray}
\end{enumerate}
$$W_t^n := \frac{w_t (H_{1:t}^n)}{\sum_{i = 1}^N w_t(H_{1:t}^i)}$$
\end{algorithm}

One disadvantage of this approach is that the the SMC approximation to $p_\theta(x_t|(\mathbf{y}_{1:T})$ deteriorates when $T-t$ is too large. {Andrieu et al. \textcolor{black}{(}2010\textcolor{black}{)}} have proposed Particle PIMH  method to overcome this difficulty. This is a class of MCMC using the SMC algorithm as its component to design its multi-dimensional proposal distributions. The advantage of this method is that the PIMH sampler does not call for SMC algorithm to generate all samples which approximate $p_{\theta}(x_{1:T}|\mathbf{y}_{1:T})$ but only to choose a sample which can be approximated for $p_\theta (x_{1:T}|\mathbf{y}_{1:T})$ (see {\textcolor{ceruleanblue}{Andrieu et al., 2010\textcolor{black}{)}}}. Algorithm \ref{PIMH algorithm} presents the PIMH method applied in our model. \\
\begin{algorithm}
\caption{PIMH algorithm} \label{PIMH algorithm}
\textsl{•} Set $ k = 0$ \\
\quad \textit{Sample $\mathbb{S} \sim p_{\theta} (h_{1:T}|(z_{1:T}, D_{1:T})) $ by SMC Algorithm \ref{SMC algorithm}},\\
\quad \textit{Draw $H_{1:T} (0) \sim \widehat{p}_{\theta} (.|(z_{1:T}, D_{1:T}))$ from $\mathbb{S}$} \\
\quad \textit {Set $\widehat{p}_{\theta} (z_{1:T}, D_{1:T})(0) = \widehat{p}_{\theta} (.|(z_{1:T}, D_{1:T}))$} \\
\textsl{•} For $ k = 1:N$\\
\begin{enumerate}
\item[(1)] \textit{Sample $\mathbb{S} \sim p_{\theta} (h_{1:T}|(z_{1:T}, D_{1:T})) $ by SMC Algorithm \ref{SMC algorithm}} \\
\textit{Draw $H_{1:T}^* \sim \widehat{p}_{\theta} (.|(z_{1:T}, D_{1:T}))$} \\
\item[(2)]
\textit{Draw U with the uniform distribution (0, 1)} \\
\textit{If $ U < \widehat{p}_{\theta} (z_{1:T}, D_{1:T})^* / {\widehat{p}_{\theta} (z_{1:T}, D_{1:T})(k-1)}$} \\
\textit{\qquad Set $H_{1:T} (k) = H_{1:T}^*$} \\
\textit{\qquad Set $\widehat{p}_{\theta} (z_{1:T}, D_{1:T}) (k) = \widehat{p}_{\theta} (z_{1:T}, D_{1:T})^*$} \\
\textit{Else} \\
\textit{\qquad Set $H_{1:T} (k) = H_{1:T} (k-1)$}\\
\textit{\qquad Set $\widehat{p}_{\theta} (z_{1:T}, D_{1:T}) (k) = \widehat{p}_{\theta} (z_{1:T}, D_{1:T}) (k -1)$}
\end{enumerate}
\end{algorithm}

In our method, we combine Particle MCMC with the maximum likelihood method to estimate the intensity parameter vector $\theta$ for the frailty correlated model. We present the implementation steps in Algorithm \ref{Particle MCMC EM algorithm}. See \hyperlink{thesentence}{\textcolor{ceruleanblue}{Nguyen \textcolor{black}{(}2023\textcolor{black}{)}}} for further discussions about the methods.
\begin{algorithm}
\caption{Particle MCMC Expectation-Maximization algorithm} \label{Particle MCMC EM algorithm}
\textsl{•} Initialize \\
\textit{\quad Set i := 0} \\
\textit{\quad Set ${\theta}^{(0)}  = (\widehat{\kappa}, 0.05, 0.01, 1)$, where $\widehat{\kappa}$ is an estimate of $\kappa$ in the model without the hidden factors} \\
\textsl{•} Loop \\
\quad \textit{Set i := i + 1} \\
\quad \textit{Sample $H^1, \allowbreak H^2, ..., H^N$ from $p_H(.|Z, \allowbreak D, \theta^{(i-1)})$ by PIMH Algorithm \ref{PIMH algorithm}} \\
\quad \textit{Employ the maximum likelihood method to estimate parameters $\theta^{(i)}$ from Eq. (\ref{Expertopinion}) using generated samples $H^1, \allowbreak H^2, ..., H^N$} \\
\quad \textit{Exit when achieving reasonable numerical convergence of the likelihood ${\cal L}$.}\\
\end{algorithm}
%
%
\vspace{-0.5cm}
\section{Major Results}
\vspace{-0.5cm}
\bookmarksetup{bold = false,color=[rgb]{0,0,0}}
\renewcommand{\theequation}{5.\arabic{equation}}
\setcounter{equation}{0}
\vspace{-0.5cm}
\subsection{Data sample}
\vspace{-0.5cm}
The dataset used to estimate the models includes: Short-term risk-free risk (3-month Treasury bill rate) was collected from the Board of Governors of the Federal Reserve System. We use the Compustat North America data set and the Center for Research in Security Prices (CRSP) database from Wharton Research Data Services. We collect quarterly and annual accounting data for companies in the nonfinancial industry in the United States. Compustat quarterly and annual files contain information regarding both short- and long-term debt. When comparing the values of Debt in Current Liabilities and Total Current Liabilities, for short-term debts, we select the greater value. When the quarterly debt values are missing, we substitute them with the annual debt values if they are available; if they are not, they are treated as the final missing values. Additionally, we include these companies’ stock market data.
Historical default rate data is collected from Moody's database. Our default measure is in a similar way to \hyperlink{thesentence}{\textcolor{ceruleanblue}{Nguyen \textcolor{black}{(}2023\textcolor{black}{)}}}. The final dataset contains 2,432 U.S Industrial category with 424,601 firm-month observations with a total of 412 defaults (272 bankruptcies and 140 other defaults) over the period from January 1980-June 2019.

\vspace{-0.5cm}
\subsection{The choice of covariates}
\vspace{-0.5cm}

The observable firm-specific/macroeconomic covariate variables used to examine and predict the defaults for the U.S. firms including:

\begin{table}[htbp]
\centering
\caption{\bf {Observable firm-specific attributes and macroeconomic factors}}\label{Definitions}
\resizebox{\textwidth}{!}
{
\begin{tabular}{clll}
  \\[-1ex]\hline\hline\\[-1.6ex]
No &  Covariates & Definitions  & Reference  \\[0.3ex]
  \hline\\[-1.6ex]     
1 & TREASURY RATE & 3-month US Treasury bill rate & \hyperlink{thesentence}{\textcolor{ceruleanblue}{Duffie et al. \textcolor{black}{(}2007\textcolor{black}{)}}}, \hyperlink{thesentence}{\textcolor{ceruleanblue}{Duffie et al. \textcolor{black}{(}2009\textcolor{black}{)}}}, \\
& & & \hyperlink{thesentence}{\textcolor{ceruleanblue}{Duan et al. \textcolor{black}{(}2012\textcolor{black}{)}}}, \hyperlink{thesentence}{\textcolor{ceruleanblue}{Nguyen \textcolor{black}{(}2023\textcolor{black}{)}}}\\   
2& SP500&Trailing 1-year return on the S\&P 500 index & \hyperlink{thesentence}{\textcolor{ceruleanblue}{Duffie et al. \textcolor{black}{(}2007\textcolor{black}{)}}}, \hyperlink{thesentence}{\textcolor{ceruleanblue}{Duffie et al. \textcolor{black}{(}2009\textcolor{black}{)}}}, \\
& & &  \hyperlink{thesentence}{\textcolor{ceruleanblue}{Duan et al. \textcolor{black}{(}2012\textcolor{black}{)}}}, \hyperlink{thesentence}{\textcolor{ceruleanblue}{Azizpour et al. \textcolor{black}{(}2018\textcolor{black}{)}}}, \\ 
& & & \hyperlink{thesentence}{\textcolor{ceruleanblue}{Nguyen \textcolor{black}{(}2023\textcolor{black}{)}}} \\ 
3 &  D2D   & Distance to Default & \hyperlink{thesentence}{\textcolor{ceruleanblue}{Duffie et al. \textcolor{black}{(}2007\textcolor{black}{)}}}, \hyperlink{thesentence}{\textcolor{ceruleanblue}{Duffie et al. \textcolor{black}{(}2009\textcolor{black}{)}}}, \\
& & &  \hyperlink{thesentence}{\textcolor{ceruleanblue}{Duan et al. \textcolor{black}{(}2012\textcolor{black}{)}}}, \hyperlink{thesentence}{\textcolor{ceruleanblue}{Nguyen \textcolor{black}{(}2023\textcolor{black}{)}}} \\    
4 & FIRM SIZE& Logarithm of the assets & \hyperlink{thesentence}{\textcolor{ceruleanblue}{Shumway \textcolor{black}{(}2001\textcolor{black}{)}}}, \hyperlink{thesentence}{\textcolor{ceruleanblue}{Nguyen \textcolor{black}{(}2023\textcolor{black}{)}}}\\
5 &  ROA & Net income to total assets & \hyperlink{thesentence}{\textcolor{ceruleanblue}{Altman \textcolor{black}{(}1968\textcolor{black}{)}}}, \hyperlink{thesentence}{\textcolor{ceruleanblue}{Shumway \textcolor{black}{(}2001\textcolor{black}{)}}}, \\
& & &  \hyperlink{thesentence}{\textcolor{ceruleanblue}{Nguyen \textcolor{black}{(}2023\textcolor{black}{)}}} \\
6 &  LEVERAGE & Total liabilities to total assets & \hyperlink{thesentence}{\textcolor{ceruleanblue}{Ohlson, 1980}}, \hyperlink{thesentence}{\textcolor{ceruleanblue}{Zmijewski, 1984}}, \\
& & &  \hyperlink{thesentence}{\textcolor{ceruleanblue}{Nguyen \textcolor{black}{(}2023\textcolor{black}{)}}}\\
7 & FIRM RETURN & Trailing 1-year stock return & \hyperlink{thesentence}{\textcolor{ceruleanblue}{Shumway \textcolor{black}{(}2001\textcolor{black}{)}}}, \hyperlink{thesentence}{\textcolor{ceruleanblue}{Duffie et al. \textcolor{black}{(}2007\textcolor{black}{)}}}, \\
& & &  \hyperlink{thesentence}{\textcolor{ceruleanblue}{Duffie et al. \textcolor{black}{(}2009\textcolor{black}{)}}}, \hyperlink{thesentence}{\textcolor{ceruleanblue}{Bharath and Shumway \textcolor{black}{(}2008\textcolor{black}{)}}}, \\
& & & \hyperlink{thesentence}{\textcolor{ceruleanblue}{Nguyen \textcolor{black}{(}2023\textcolor{black}{)}}} \\
\hline \hline
\end{tabular}
}
  \begin{minipage}{0.97\textwidth}
{\footnotesize  \textit{Notes}: The details of observable covariates used to examine and predict the frailty correlated default model with prior distributions.
    \par}
\end{minipage}
\end{table}

\begin{enumerate}
\item[1.] \text{The 3-month Treasury bill rate} (TREASURY RATE): The 3-Month Treasury bill is a short-term U.S. government security with a constant 3-month maturity. The Federal Reserve computes yields for constant maturities by interpolating points along a Treasury yield curve comprised of actively traded issues with term maturities. It is a risk-free rate and has a significant impact on monetary policy (see, for example, \hyperlink{thesentence}{\textcolor{ceruleanblue}{Duffie et al., 2007, 2009}}; \hyperlink{thesentence}{\textcolor{ceruleanblue}{Duan et al., 2012}}; \hyperlink{thesentence}{\textcolor{ceruleanblue}{Azizpour et al., 2018}}; \hyperlink{thesentence}{\textcolor{ceruleanblue}{Nguyen, 2023}}).
\item[2.]\text{The Trailing 1-year return on the S$\&$P 500} (SP 500): This variable measures the market return, and its importance has been documented in previous studies (see, for example, \hyperlink{thesentence}{\textcolor{ceruleanblue}{Duffie et al., 2007, 2009}}; \hyperlink{thesentence}{\textcolor{ceruleanblue}{Duan et al., 2012}}; \hyperlink{thesentence}{\textcolor{ceruleanblue}{Azizpour et al., 2018}}; \hyperlink{thesentence}{\textcolor{ceruleanblue}{Nguyen, 2023}}).
\item[3.] Distance to Default (D2D): This variable is defined as the number of standard deviations of the annual asset growth of a firm where the firm's assets are higher than its liabilities (see {\textcolor{ceruleanblue}{Merton \textcolor{black}{(}1974\textcolor{black}{)}}} for further discussion on this variable). We construct this variable in a similar way to
{\textcolor{ceruleanblue}{Vassalou and Xing  \textcolor{black}{(}2004\textcolor{black}{)}}}, \label{Vassalou and Xing (2004)} \hyperlink{thesentence}{\textcolor{ceruleanblue}{Hillegeist et al. \textcolor{black}{(}2004\textcolor{black}{)}}}, \label{Hillegeist et al. (2004)} \hyperlink{thesentence} {\textcolor{ceruleanblue}{Bharath and Shumway \textcolor{black}{(}2008\textcolor{black}{)}}}, and \hyperlink{thesentence} {\textcolor{ceruleanblue}{Nguyen \textcolor{black}{(}2023\textcolor{black}{)}}}. {\textcolor{ceruleanblue}{Duffie et al. \textcolor{black}{(}2007, 2009\textcolor{black}{)}}} and {\textcolor{ceruleanblue}{Nguyen \textcolor{black}{(}2023\textcolor{black}{)}}} find a negative and significant relationship between distance to default and the default intensity of the US firms in Industry category. {\textcolor{ceruleanblue}{Duan et al. \textcolor{black}{(}2012\textcolor{black}{)}}} also show that the default risk of U.S industry and financial firm firms is significantly and negatively associated with the Distance-to-Default variable.
\item[4.] Firm size (FIRM SIZE): This variable is used to show the measure or quantity of a company's assets. The importance of this variable was documented in a study by  \hyperlink{thesentence}{\textcolor{ceruleanblue}{Shumway \textcolor{black}{(}2001\textcolor{black}{)}}}, \hyperlink{thesentence}{\textcolor{ceruleanblue}{Duan et al. \textcolor{black}{(}2012\textcolor{black}{)}}}, and \hyperlink{thesentence}{\textcolor{ceruleanblue}{Nguyen \textcolor{black}{(}2023\textcolor{black}{)}}}. Firm size is calculated as the logarithm of the assets.
\item[5.] Return on assets ratio (ROA): This is a financial ratio that indicates a company's ability to generate profit relative to the value of its assets. A higher ROA expressed as a percentage indicates that a company can generate more profits from its assets. A lower ROA indicates productivity and the company's ability to better its balance sheet management. The return on assets ratio is computed as a ratio of net income to total assets. In the default literature, the profitability ratio is a traditional variable, and its importance has been pointed out since \hyperlink{thesentence}{\textcolor{ceruleanblue}{Altman \textcolor{black}{(}1968\textcolor{black}{)}}} and is widely used in the finance literature, such as \hyperlink{thesentence}{\textcolor{ceruleanblue}{Shumway \textcolor{black}{(}2001\textcolor{black}{)}}}, \hyperlink{thesentence} {\textcolor{ceruleanblue}{Duan et al. \textcolor{black}{(}2012\textcolor{black}{)}}}, {\textcolor{ceruleanblue}{Nguyen \textcolor{black}{(}2023\textcolor{black}{)}}}.
\item[6.] \text{Financial leverage ratio} (LEVERAGE): This ratio, also known as the debt ratio, is used to assess a company's ability to meet its long-term (one year or longer) debt obligations. These obligations consist of interest payments, the ultimate principal payment, and any other fixed obligations, such as lease payments. This ratio is calculated as the ratio of total liabilities to total assets (see, for example, \hyperlink{thesentence}{\textcolor{ceruleanblue}{Ohlson, 1980}};\label{Ohlson (1980)} \hyperlink{thesentence}{\textcolor{ceruleanblue}{Zmijewski, 1984}}; \label{Zmijewski(1984)} \hyperlink{thesentence}{\textcolor{ceruleanblue}{Nguyen, 2023}}).
\item[7.] Trailing 1-year firm stock return (FIRM RETURN): This variable is suggested by \hyperlink{thesentence}{\textcolor{ceruleanblue}{Shumway \textcolor{black}{(}2001\textcolor{black}{)}}} \label{Shumway (2001)} and is widely used in the finance literature (see, for example, \hyperlink{thesentence}{\textcolor{ceruleanblue}{Bharath and Shumway, 2008}}; \label{Bharath and Shumway (2008)} \hyperlink{thesentence}{\textcolor{ceruleanblue}{Duffie et al., 2007, 2009}}; \label{Duffie et al. (2007, 2009)} \hyperlink{thesentence}{\textcolor{ceruleanblue}{Nguyen, 2023}}). We use a similar formula as \hyperlink{thesentence}{\textcolor{ceruleanblue}{Shumway \textcolor{black}{(}2001\textcolor{black}{)}}} \label{Shumway (2001)} and \hyperlink{thesentence}{\textcolor{ceruleanblue}{Nguyen \textcolor{black}{(}2023\textcolor{black}{)}}} to compute this variable. 
\end{enumerate}

Table \ref{Definitions} and Table \ref{Summary statistics}
provide definitions and summary statistics for all research covariates used in the sample to predict the frailty-correlated default models with different prior distributions.

\begin{table}[htbp]
\centering
\caption{\bf {Summary statistics of observable firm-specific attributes and macroeconomic factors}}\label{Summary statistics}
\resizebox{\textwidth}{!}
{
\begin{tabular}{clccccccc}
  \\[-1ex]\hline\hline\\[-1.6ex]
& Variable & Mean  & SD   &  Minimum & Median & Maximum \\[0.3ex]
  \hline\\[-1.6ex]     
  &\textbf {Macroeconomics covariates}\\
    & TREASURY RATE   & 4.6837 & 3.1343 & 0.0100  &  4.9500  &  16.300 \\[0.5ex]
    & SP500 & 0.1048 & 0.1534 & -0.5542 &  0.1225 &  0.4452 \\[0.5ex]
 
&\textbf {Firm-specific covariates}  \\    
    & \textbf {D2D}  \\          
    & Defaults       & 0.0325    & 1.3529  & -14.1997  & 0.0332   & 4.9388 \\[0.5ex]
    & Nondefaults    & 1.9052    & 1.4412  & -5.4534   & 1.8025   & 48.6861 \\[0.5ex]
    & \textbf {FIRM SIZE}  \\          
    & Defaults       & 20.1936 & 1.5811 & 15.1688 & 20.1070  & 26.3362     \\[0.5ex]
    & Nondefaults    & 21.4783 & 1.8422 & 13.5392 & 21.4586  & 27.9370 \\[0.5ex]
    & \textbf {ROA}  \\          
    & Defaults       & -0.0096 & 0.1026 & -5.3156 & 0.0036   & 4.1160 \\[0.5ex]
    & Nondefaults    & 0.0105  & 0.0449 & -3.5341 & 0.0128   & 2.9270\\[0.5ex]
    & \textbf {LEVERAGE}  \\          
    & Defaults       & 0.7060  & 0.3792 &  0.0000 & 0.6662   & 7.6641  \\[0.5ex]
    & Nondefaults    & 0.5671  & 0.2438 &  0.0000 & 0.5588   & 8.2774 \\[0.5ex]
    & \textbf {FIRM RETURN}  \\          
    & Defaults       & -0.0339 & 1.3053 & -2.8998 & -0.0962  & 45.8583 \\[0.5ex]
    & Nondefaults    & -0.0356 & 0.4603 & -2.9045 & -0.0444  & 5.4282 \\[0.5ex]
\hline \hline
\end{tabular}
}
  \begin{minipage}{0.97\textwidth}
{\footnotesize  \textit{Notes}: The historical default rates comprises 424,601 month observations between January 1980 and June 2019.  
    \par}
\end{minipage}
\end{table}
\vspace{-0.5cm}
\subsection{Parameter estimates}
\vspace{-0.5cm}
We estimate both default models with both uniform and subjective prior distribution, which enables us to compare two models easily. Table \ref{uniform prior} shows the estimates for parameters of default intensities with a uniform prior distribution.
\begin{table}[htbp]
\centering
\caption{\bf {Estimation results of default intensity with non-informative prior distribution}}\label{uniform prior}
\resizebox{\textwidth}{!}
{
\begin{tabular}{lcccccc}
  \\[-1ex]\hline\hline\\[-1.6ex]
 Predictor & \multirow{2}{*}{\text{ Coefficient }}  & Asymptotic   &  \multirow{2}{*}{\text{t-Statistic}} & \multicolumn{2}{c}{\text{95\% Confidence Interval }} \\[0.3ex]
  \cline{5-6}\\[-1.6ex]
                               &          &     Standard Error    &       &  Lower Bound &  Upper Bound \\[0.5ex]
  \hline\\[-1.6ex] 
\textbf{Macroeconomics covariates}:\\                             
Constant                        & -3.1263   & 0.7673   & -4.07   &  -4.6303     &  -1.6223 \\[0.5ex]
TREASURY RATE     & -0.1231   & 0.0231   & -5.33   &  -0.1685     &  -0.0777 \\[0.5ex]
SP500 & -0.9093   & 0.2832   & -3.21   &  -1.4645     &  -0.3540\\[0.5ex]
\textbf{Firm-specific covariates}:\\ 
D2D            & -0.6099   & 0.0202   & -30.19  &  -0.6496     &  -0.5703\\[0.5ex]
FIRM SIZE                     & -0.1838   & 0.0355   & -5.18   &  -0.2535     &  -0.1142 \\[0.5ex]
ROA        & -0.3691   & 0.0941   & -3.92   &  -0.5536     &  -0.1846 \\[0.5ex]
LEVERAGE      & 0.5293    & 0.0462   & 11.46   &   0.4387     &  0.6200 \\[0.5ex]
FIRM RETURN   & -1.1282   & 0.0825   & -13.67  &  -1.2900     &  -0.9663 \\[0.5ex]
\textbf{Frailty effect}:\\ 
Hidden-factor volatility        & 0.1096    & 0.0061   & 17.97   &  0.0975      & 0.1216  \\[0.5ex]
Hidden-factor mean reversion    & 0.4360    & 0.0546   & 7.98    &  0.3288      & 0.5432 \\[0.5ex] 
Brownian motion volatility      & 8.4610    &0.3394    & 24.93   &  7.7956      & 9.1264  \\[0.5ex]        
  \hline 
No. of firm-month observations & 424,601 \\
Log-likelihood & -2379.61\\
\hline \hline
\end{tabular}
}
  \begin{minipage}{0.97\textwidth}
{\footnotesize  \textit{Notes}: Asymptotic standard errors of the estimated parameters are computed using the Hessian matrix. 
    \par}
\end{minipage}
\end{table}

\begin{table}[htbp]
\centering
\caption{\bf {Estimates of the frailty correlated model with subjective prior distribution}}\label{subjective prior}
\resizebox{\textwidth}{!}
{
\begin{tabular}{lcccccc}
  \\[-1ex]\hline\hline\\[-1.6ex]
 Predictor & \multirow{2}{*}{\text{ Coefficient }}  & Asymptotic   &  \multirow{2}{*}{\text{t-Statistic}} & \multicolumn{2}{c}{\text{95\% Confidence Interval }} \\[0.3ex]
  \cline{5-6}\\[-1.6ex]
                               &          &     Standard Error    &       &  Lower Bound &  Upper Bound \\[0.5ex]
  \hline\\[-1.6ex]              
\textbf{Macroeconomics covariates}:\\              
Constant                           & -3.4556  &0.6259  & -5.52  & -4.6825  & -2.2288 \\[0.5ex]
TREASURY      & -0.1175  &0.0184  & -6.37  & -0.1536  & -0.0813 \\[0.5ex]
SP500    & -1.0620  &0.2306  & -4.60  & -1.5142  & -0.6099  \\[0.5ex]
\textbf{Firm-specific covariates}:\\ 
D2D              & -0.6309  &0.0169  & -37.30 & -0.6641  & -0.5977 \\[0.5ex]
FIRM SIZE                         & -0.1856  &0.0289  & -6.42  & -0.2422  & -0.1290 \\[0.5ex]
ROA          & -0.3807  &0.0780  &-4.8792   & -0.5336  & -0.2277\\[0.5ex]
LEVERAGE          & 0.5570   &0.0380  & 14.6277  & 0.4823   &  0.6316 \\[0.5ex]
FIRM RETURN     & -1.2134  &0.0676  & -17.93 & -1.3461  & -1.0808 \\[0.5ex]
\textbf{Frailty effect}:\\ 
Hidden-factor volatility           & 0.0897   &0.0040  &22.00   & 0.0817   & 0.0977  \\[0.5ex]
Hidden-factor mean reversion       & 0.6189   &0.0590  &10.47    & 0.5031   & 0.7347 \\[0.5ex]  
Brownian motion volatility         &12.5069   &0.4354  &28.7216   & 11.6534  & 13.3604 \\[0.5ex]       
  \hline 
No. of firm-month observations & 424,601 \\
Log likelihood        & -2202.45\\
\hline \hline
\end{tabular}
}
  \begin{minipage}{0.97\textwidth}
{\footnotesize  \textit{Notes}:  Table reports the estimation result of the frailty correlated model combined with subjective prior distribution. Asymptotic standard errors of the estimated parameters are calculated by the Hessian matrix. Given $\mu$ and $\Sigma$ below.
    \par}
\end{minipage}
\end{table}
 \small  
\begin{landscape}
\begin{align*} \label{Covariance mt}
 \mu &= (-3.1  \quad -0.6  \quad   -1.1  \quad     -0.1  \quad     -0.9  \quad     -0.18  \quad   -0.36 \quad   0.53 \quad    0.1)  \nonumber\\
\Sigma &=
\begin{bmatrix}
0.540000  & -0.004164 &  -0.008542 &  0.006700 &  -0.017213 &  0.024840 & -0.008855 & 
0.005788  & -0.000056 \\
-0.004164 &  0.000440 &   0.000327 &  -0.000088&  0.000446  &  0.000206 &  -0.000054 &
-0.000067 &  0.000084 \\
-0.008542 &  0.000327 &   0.006385 &  0.000111 &  0.000912  &  0.000272 & -0.000554  &
-0.000534 & -0.000018\\
0.006700  & -0.000088 &   0.000111 &  0.000472 &  0.002533  & -0.000251 &  0.000129 &
-0.000091 &  0.000023 \\
-0.017213 &  0.000446 &   0.000912 &  0.002533 &  0.074100  &  0.000619 &  -0.001904 &
-0.000771 &  0.000015 \\	
0.024840  &  0.000206 &   0.000272 & -0.000251 &  0.000619  &  0.001164 &  0.000457 &       -0.000189 &  0.000041 \\
-0.008855 & -0.000022 &  -0.000554 & 0.000129  & -0.001904  &  0.000457 &  0.008680 &        -0.002102 &  0.000045\\
0.005788  & -0.000053 &  -0.000534 & -0.000091 & -0.000771  & -0.000189 & -0.002102 &            0.002021  &  0.000010 \\
-0.000045 &  0.000041 &  -0.000019 &  0.000026 & -0.000043  & 0.000021  &  0.000057 &            0.000009  &  0.000023 
    \end{bmatrix} 
\end{align*}
\end{landscape}
\normalsize

From Table \ref{subjective prior}, it can be seen that all these variables are statistically significant at traditional confidence levels. The estimate of Distance to Default of $-0.6309$  indicates that a negative shock to the distance to default by one standard deviation increases the default intensity by $\approx 87.91 \%$. Among firm-specific variables, Distance to Default, which is the volatility-adjusted leverage measure, shows its dominant role in explaining a significant variation of the default intensity. 
The result of Firm size indicates that
larger firms often have more financial flexibility than smaller firms, which can help them better overcome financial distress. The coefficient of Return on assets ratio confirms that firms with high-profits relative to assets are less likely to go bankrupt. The result of financial leverage ratio reports that the higher the debt ratios, the higher the default risk of firms. The 1-year trailing stock return covariate is  statistically significant and negatively related to the default intensities of the firms. 
The observable macroeconomic variables chosen in this study are highly economically and statistically significantly negatively associated with the default intensities of the firms. 

The role of the frailty effect is not relatively large in our dataset. The volatility and the mean reversion of the hidden factor, which determine the dependence of the unobserved default intensities on the latent variable $H_t$, have a highly economically and statistically significant impact on the default intensities of the firms. The frailty volatility is the coefficient $\xi$ of the dependence of the default intensity on the OU frailty process $H$. The coefficient of $0.1096$ shows us that an increase of $1\%$ of the latent factor volatility will increase the unobserved default intensities by $10.96\%$ monthly. This finding is consistent with  \hyperlink{thesentence}{\textcolor{ceruleanblue}{Duffie et al. \textcolor{black}{(}2009\textcolor{black}{)}}} and \hyperlink{thesentence}{\textcolor{ceruleanblue}{Nguyen \textcolor{black}{(}2023\textcolor{black}{)}}}. The estimated mean reversion  $\eta$  of frailty factor is approximately $43.60\%$ monthly. Brownian motion volatility is statistically significantly positive.
In general, signs of coefficients in the frailty correlated defaults models are no surprise. It can be seen from Tables \ref{uniform prior} and \ref{subjective prior} that the signs and scales of estimates in both cases where models with uniform and subjective prior distributions are similar. 

\vspace{-0.5cm}
\section{Out-of-sample performance and robustness check}
\vspace{-0.5cm}
To evaluate the model performance, we use the cumulative accuracy profile (CAP) and the accuracy ratio (AR). The companies are divided into two equal groups: estimation and evaluation. We estimate the parameters based on the estimation group and then evaluate the prediction accuracy using the evaluation group. The implementation steps are shown as follows: Firstly, we estimate parameters in the frailty correlated default model with subjective prior distribution using the historical default rates in the period from 1981 to 2011. Secondly, using the estimation results obtained from Step 1, we forecast the data for the period from 2012 to 2018 based on the covariates time series model for observable firm-specific/macroeconomic covariates. Thirdly, we forecast the data of the frailty variable for the period (2012-2018) using the PIMH Algorithm \ref{PIMH algorithm}. Fourthly, after obtaining the estimates from Step 1 and the data obtained from Steps 2 and 3, we compute the default probability based on Eq. \ref{default probability}. Lastly, we can determine a CAP and its associated AR. The CAPs and ARs for the out-of-sample prediction horizons are displayed in \text{\textcolor{palecarmine}{Figure 1}} and \text{\textcolor{palecarmine}{Figure 2}}.

\begin{figure} [H]
\centering
\includegraphics[width=100mm]{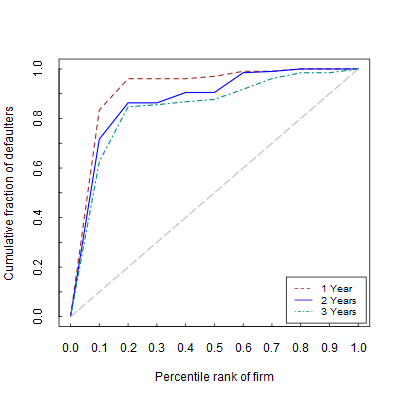}
\caption{This graph illustrates the out-of-sample cumulative accuracy profiles (power curves) over the entire sample period (1980-2019) for various prediction horizons. The companies are divided into two equal groups: estimation and evaluation. We estimate the parameters based on the estimation group and then evaluate the prediction accuracy using the evaluation group.
The power curve illustrates 20\% of companies with the most capacity of default over the different horizons in the frailty correlated default model with subjective prior.} \label{CurveFor123Year}
\end{figure}

\begin{figure}  [H]
\centering
\includegraphics[width=100mm]{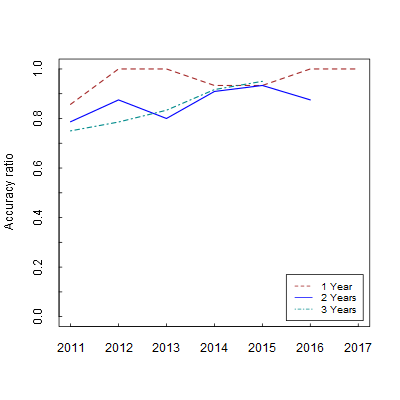}
\caption{This graph illustrate average accuracy ratios for out-of-sample prediction in three prediction accuracy horizons for frailty correlated default models with expert opinion} \label{AR} 
\end{figure}

\begin{table}[htbp]
\centering
\caption{\bf {Prediction accuracy for frailty correlated default model with different prior distribution}}\label{Accuracy ratios}
\resizebox{\textwidth}{!}
{
\begin{tabular}{lcccccccccc}
  \\[-1ex]\hline\hline\\[-1.6ex]
           & T (Year)  & 2012   & 2013   & 2014   & 2015  & 2016  & 2017 & 2018  &Average\\[0.3ex]
  \hline\\[-1.6ex]                         
                   & 1 & 85.71 & 100.00 & 100.00 & 87.50 & 91.67 & 100.00 & 100.00 & 95.00 \\[0.5ex]
\textbf{Uniform prior}   & 2 &       & 77.78  & 83.33  & 80.00 & 89.47 & 93.33  & 87.50  & 85.23 \\[0.5ex]
                   & 3 &       &        & 72.73  & 78.57 & 78.68 & 90.91  & 95.00  & 83.18 \\[0.5ex]     
        \hline     
                           & 1 & 85.71 & 100.00 & 100.00 & 93.33 & 93.33 & 100.00 & 100.00 & 96.05 \\[0.5ex]
\textbf{Subjective prior}& 2 &       &  78.68 & 87.50  & 80.00 & 90.91 & 93.33  & 87.50  & 86.32 \\[0.5ex]
                   & 3 &       &        & 75.00  & 78.57  & 83.33  & 91.67 & 95.00 & 84.71 \\[0.5ex]
         
\hline \hline
\end{tabular}
}
  \begin{minipage}{0.97\textwidth}
{\footnotesize  \textit{Notes}: The table reports the accuracy ratios for the out-of-sample prediction for different prediction horizons. In particular, individual accuracy ratios and average accuracy ratios for the model with uniform and subjective prior distributions over three default prediction horizons (2012-2018, 2013-2018, and 2014-2018) are presented.
    \par}
\end{minipage}
\end{table}

Table \ref{Accuracy ratios} reports the results of out-of-sample predictions of frailty correlated models with uniform and subjective prior distributions. From two default models, it can be seen that the prediction ratios of the frailty correlated default model with subjective prior distribution are higher than those of the model with uniform prior distribution. The out-of-sample prediction accuracy for 1-year prediction on average is good. Specifically, 95 percent for the frailty correlated model with a non-informative prior distribution and 96.05 percent for the model with a subjective prior distribution). When the time horizon for predictions is extended to three years, the AR of the models suffers a significant decline, falling to 85.23 percent for frailty-correlated models with uniform prior distributions and 86.32 percent for those with subjective prior distributions. We also perform out-of-sample default predictions using the logistic regression method \footnote{I would like to thank a referee for suggesting this}
 to compare the accuracy with our proposed method in Table \ref{Logistic Regression}. The results show that our method has better prediction power compared with the logistic regression method.
\begin{table}[htbp]

\centering
\caption{\bf Compare default prediction accuracy between the Logistic Regression method and the Particle MCMC Expectation-Maximization method}\label{Logistic Regression}
\resizebox{\textwidth}{!}
{
\begin{tabular}{lcccccccccc}
  \\[-1ex]\hline\hline\\[-1.6ex]
           & T (Year)  & 2012   & 2013   & 2014   & 2015  & 2016  & 2017 & 2018  &Average\\[0.3ex]
  \hline\\[-1.6ex]                         
                   & 1 & 80.00 & 75.00 & 50.00 & 80.00 & 90.00 & 100.00 & 100.00 & 82.14 \\[0.5ex]
\textbf{Logistic Regression}   & 2 &       & 66.67  & 66.67  & 57.14 & 78.57 & 90.91  & 80.00  & 73.33 \\[0.5ex]
                   & 3 &       &        & 62.50  & 54.55 & 64.28 & 80.00  & 93.33  & 70.93 \\[0.5ex]   
        \hline \\     
\textbf{Particle MCMC}  & 1 & 85.71 & 100.00 & 100.00 & 93.33 & 93.33 & 100.00 & 100.00 & 96.05 \\[0.5ex]
\textbf{Expectation-Maximization}& 2 &       &  78.68 & 87.50  & 80.00 & 90.91 & 93.33  & 87.50  & 86.32 \\[0.5ex]
\textbf{with subjective prior}  & 3 &       &        & 75.00  & 78.57  & 83.33  & 91.67 & 95.00 & 84.71 \\[0.5ex]
\hline \hline
\end{tabular}
}
  \begin{minipage}{0.97\textwidth}
{\footnotesize \textit{Notes}: The table reports the accuracy ratios for the out-of-sample default prediction in 1-year, 2-years and 3-years using the Logistic Regression method and Particle MCMC Expectation-Maximization with subjective prior method.
    \par}
\end{minipage}
\end{table}
 
Overall, two notable conclusions can be drawn from these parameter estimation results: (i) The 1-year prediction for both models is good and when the prediction horizons increase, the prediction accuracy of the models decreases significantly. (ii) It can be seen that there has not been much difference about prediction accuracy ratios for frailty correlated default models with non-informative and subjective prior distributions over three out-of-sample prediction horizons, including 2012-2018 for 1-year default distribution, 2013-2018 for 2-year default prediction, and 2014-2018 for 3-year default prediction. 

To check the robustness of the estimation results for the frailty-correlated default model with subjective prior distribution, we estimated a subperiod from 1980 to 2011 as a sensitivity test. The outcomes correspond with the signs and magnitude of the entire sample. On the other hand, the value of log-likelihood in the frailty correlated default model with subjective prior distribution (-2202.45) is larger than that in the frailty correlated default model with non-informative prior distribution (-2379.61), which confirms that the frailty correlated default model should incorporate the expert opinion.
\vspace{-0.5cm}
\section{Concluding remarks and limitations}
\vspace{-0.5cm}
Risk assessment is part of the decision-making process in many fields of discipline including finance. In the financial distress literature, the credit risk evaluation entails the evaluation of the hazard of potential future exposure or probable loss to lenders in the context of lending activities. The effective management of credit risk is a crucial aspect of risk management and crucial to the long-term survival of any bank. Credit risk management's objective is to maximise the bank's risk-adjusted return by keeping credit risk exposure within acceptable limits. The ability to accurately forecast a company's financial distress is a major concern for many stakeholders. This practical relevance has motivated numerous studies on the topic of predicting corporate financial distress. To improve the prediction accuracy of default models, the utilization of expert judgement in the decision-making process is common in practice as there may not be enough a statistically significant amount of empirical evidence to reliably estimate parameters of complicated models. This problems is considered to be of central interest of simulating a number of debates in the empirical literature regarding the issue of Bayesian inference. 

This paper proposes a method to add expert judgement to the frailty correlated default risk model in \hyperlink{thesentence}{\textcolor{ceruleanblue}{Duffie et al. \textcolor{black}{(}2009\textcolor{black}{)}}} by incorporating subjective prior distributions into the model. Then we employ the Bayesian method coupled with a Particle MCMC approach in \hyperlink{thesentence}{\textcolor{ceruleanblue}{Nguyen\textcolor{black}{(} 2023\textcolor{black}{)}}} in order to evaluate the unknown parameters and predict the default risk models on a historical defaults dataset of 424,601 firm-month observations from January 1980 to June 2019 of 2,432 U.S. industrial firms. We compare the prediction results of the frailty correlated default risk model with uniform and subjective prior distributions together.
The findings show that the 1-year prediction for both models are pretty good and the prediction accuracy of models decrease considerably as the prediction horizons increase. The results also indicate that prediction accuracy ratios for frailty-correlated default models with non-informative and subjective prior distributions over various prediction horizons are not significantly different. Specifically, the out-of-sample prediction accuracy for the frailty correlated default models with uniform distribution is slightly higher than that of the frailty correlated default models with informative prior distribution over three out-of-sample prediction horizons, including 2012-2018 for 1-year default distribution, 2013-2018 for 2-year default prediction, and 2014-2018 for 3-year default prediction.

The frailty correlated default model with expert opinion has been designed to estimate and predict the default risk of corporations. The model is adaptable to accommodate any context. However, the model also has its limitations. Firstly, one of the main limitation is that we cannot access inputs of data for expert opinion; therefore, to some certain extent, our results also depend on how we assume the values of priors. Accordingly, the prediction accuracy can be slightly different.
It is observed that Bayes factors are sensitive to the selection of unknown parameters from informative prior distributions, which has a greater likelihood of influencing the posterior distribution. As a consequence, it generates debates regarding the selection of priors. According to \hyperlink{thesentence}{\textcolor{ceruleanblue}{Kass and Raftery \textcolor{black}{(}1995\textcolor{black}{)}}}, non-informative priors may also contribute to posterior estimate instability and convergence of the sampler algorithm. Choosing informative priors and establishing a connection with expert opinion are still the subject of debate in academic research, and interesting stories about them are still being continued. Therefore, future work should use an actual data of expert opinion, which may be feasibly conducted in the age of big data. Recently, there have been research on the default prediction combined with expert opinion using machine learnings, such as \hyperlink{thesentence}{\textcolor{ceruleanblue}{Lin and McClean \textcolor{black}{(}2001\textcolor{black}{)}}}, \hyperlink{thesentence}{\textcolor{ceruleanblue}{Kim and Han \textcolor{black}{(}2003\textcolor{black}{)}}}, 
{\textcolor{ceruleanblue}{Zhou et al. \textcolor{black}{(}2015\textcolor{black}{)}}, and {\textcolor{ceruleanblue}{Gepp et al. \textcolor{black}{(}2018\textcolor{black}{)}}}. However, these studies adopt machine learning techniques with single classifiers and observable variables. Future work can adopt a meta-learning framework to examine and predict defaults with expert opinion at the firm level.
\vspace{-0.5cm}
\subsection*{Acknowledgments}
\vspace{-0.5cm}
I would like to thank Professor Tak Kuen Siu and Professor Tom Smith for their insightful comments and suggestions. I would also like to thank the referees for their helpful comments and suggestions.

\clearpage

\end{document}